\documentclass[aps,showpacs,twocolumn,floatfix]{revtex4}

\usepackage{amsmath,amssymb}
\usepackage{epsfig}
\usepackage{graphicx}
\usepackage{bm}
\usepackage{amsmath}
\newcommand{\beq}{\begin{equation}}
\newcommand{\eeq}{\end{equation}}
\newcommand{\beqa}{\begin{eqnarray}}
\newcommand{\eeqa}{\end{eqnarray}}
\newcommand{\beqar}{\begin{eqnarray*}}
\newcommand{\eeqar}{\end{eqnarray*}}

\newcommand\munu{\ensuremath{{\mu\nu}}}

\begin{document}

\title{Quasiblack holes with pressure: relativistic charged
spheres as the frozen stars}

\author{Jos\'e P. S. Lemos}
\affiliation{Centro Multidisciplinar de Astrof\'{\i}sica -- CENTRA,
Departamento de F\'{\i}sica,  Instituto Superior T\'ecnico - IST,
Universidade T\'ecnica de Lisboa - UTL,
Av. Rovisco Pais 1, \\ 1049-001 Lisboa,  Portugal, \\\&\\
Institute of Theoretical Physics - ITP, Freie Universit\"at Berlin,
Arnimallee 14, D-14195 Berlin, Germany,
email: joselemos@ist.utl.pt}
\author{Vilson T. Zanchin}
\affiliation{Centro de Ci\^encias Naturais e Humanas, Universidade
Federal do ABC, Rua Santa Ad\'elia 166, 09210-170 Santo Andr\'e, SP,
Brazil,\\\&\\
Coordenadoria de Astronomia e Astrof\'{\i}sica, Observat\'orio
Nacional-MCT, Rua General Jos\'e Cristino 77, 20921-400 Rio de Janeiro,
Brazil,
email: zanchin@ufabc.edu.br}

\begin{abstract}
{In general relativity coupled to Maxwell's electromagnetism and
charged matter, when the gravitational potential $W^2$ and the
electric potential field $\phi$ obey a relation of the form $W^{2}=
a\left(-\epsilon\, \phi+ b\right)^2 +c$, where $a$, $b$ and $c$ are
arbitrary constants, and $\epsilon=\pm1$ (the speed of light $c$ and
Newton's constant $G$ are put to one), a class of very interesting
electrically charged systems with pressure arises. We call the
relation above between $W$ and $\phi$, the Weyl-Guilfoyle relation,
and it generalizes the usual Weyl relation, for which $a=1$. For both,
Weyl and Weyl-Guilfoyle relations, the electrically charged fluid, if
present, may have nonzero pressure. Fluids obeying the Weyl-Guilfoyle
relation are called Weyl-Guilfoyle fluids.  These fluids, under the
assumption of spherical symmetry, exhibit solutions which can be
matched to the electrovacuum Reissner-Nordstr\"om spacetime to yield
global asymptotically flat cold charged stars. We show that a
particular spherically symmetric class of stars found by Guilfoyle has
a well-behaved limit which corresponds to an extremal
Reissner-Nordstr\"om quasiblack hole with pressure, i.e., in which the
fluid inside the quasihorizon has electric charge and pressure, and the
geometry outside the quasihorizon is given by the extremal
Reissner-Nordstr\"om metric. The
main physical properties of such charged stars and quasiblack holes
with pressure are analyzed.  An important development provided by
these stars and quasiblack holes is that without pressure the
solutions, Majumdar-Papapetrou solutions, are unstable to kinetic
perturbations. Solutions with pressure may avoid this instability. If
stable, these cold quasiblack holes with pressure, i.e.,
these compact relativistic charged spheres, are really frozen
stars.}


\end{abstract}


\maketitle

\section {Introduction}
\label{sec-introd}

Frozen star was a name advocated in the 1960s by Zel'dovich
\cite{zeldovichbookorpaper} to give to an object which is now called,
after Wheeler's suggestion in 1968, a black hole, see e.g.,
\cite{mtw}, see also \cite{israel} for the historical evolution of the
concept. As we will see, a frozen star is actually a quasiblack hole.

Quasiblack holes are objects on the verge of becoming black holes but
avoid it, their boundary approaches their own gravitational radius as
closely as one likes. They appeared in Einstein-Maxwell systems with
special matter, as had been found not so explicitly in \cite{bonnor1}
(see also \cite{bonnor2}) and then thoroughly discussed in
\cite{lemosweinberg,kleber,lemoszanchin1,lemoszanchin2008}.  They also
arise, and indeed were first discussed as such, in the context of
Einstein---Yang-Mills--Higgs systems \cite{lue1,lue2}.  These objects
have well-defined properties \cite{lz1,lz2,lz3,lz4,lz5}; for instance,
in general, regular quasiblack holes are extremal.  Rotating
quasiblack holes are also known, see \cite{bardeenwag} with hindsight
and \cite{meinel,lemoszaslavskii} for the properties of such objects.

There are several ways of solving Einstein's equations when the
gravitational field is coupled to matter. Since we will focus on
Einstein-Maxwell static systems with matter we can perhaps distinguish
two ways.  One we can call Weyl's way, the other is the
Tolman-Oppenheimer-Volkoff (TOV) way.

Let us deal with Weyl's way now.  Weyl \cite{weyl1}, while studying
stationary electric fields in vacuum Einstein-Maxwell theory,
perceived that it is interesting to consider a functional relation
between the metric potential $g_{tt}\equiv W^2(x^i)$ and the electric
potential $\phi(x^i)$ (where $x^i$ represent the spatial coordinates,
$i=1,2,3$) given by the ansatz $ W=W(\phi)$. By assuming the system is
vacuum and axisymmetric, Weyl found that such a relation must be
quadratic in $\phi$.  One can go beyond vacuum solutions, and consider
fluids which obey the Weyl relation, obtaining their properties.
These fluid systems were explored later by Majumdar and Papapetrou
\cite{majumdar,papapetrou} who in going beyond vacuum found
equilibrium configurations for charged extremal matter where the
electric repulsion is balanced by the gravitational attraction, and
moreover for this particular case they showed there is a perfect
square relationship between $W$ and $\phi$, see also
\cite{lemoszanchin2005} for the extension of these results to
$d$ dimensions.  An interesting development on Weyl's work was
performed much later by Guilfoyle \cite{guilfoyle} who considered
charged fluid distributions with the hypothesis that the functional
relation between the gravitational and the electric potential,
$W=W(\phi)$ is given by $W^{2}= a\left(-\epsilon\, \phi+ b\right)^2
+c$, where $a$, $b$ and $c$ are arbitrary constants, and
$\epsilon=\pm1$ (the speed of light $c$ and Newton's constant $G$ are
put to one). This relation generalizes the usual Weyl relation, for
which $a=1$, and it allows a further beautiful relationship between
the various field and matter quantities \cite{lemoszanchin2009}, which
in turn generalizes the Gautreau and Hoffman results \cite{gautreau}
for fluids obeying a pure Weyl relation.  For both, Weyl and
Weyl-Guilfoyle relations, the electrically charged fluid, if present,
may have nonzero pressure, and this turns out to be important in our
context.  Fluids obeying the Weyl-Guilfoyle relation are called
Weyl-Guilfoyle fluids.

Up to now we have only mentioned properties of the fluid itself, be it
a Weyl-Guilfoyle fluid or, in the particular case of zero pressure, a
Majumdar-Papapetrou fluid.  But matter solutions can be matched into
vacuum asymptotically flat solutions, yielding star solutions which,
besides having the local properties associated to the fluid itself,
have global properties for the spacetime as a whole.  Thus,
Majumdar-Papapetrou matter solutions, when joined to vacuum solutions
yield star solutions, the Bonnor stars \cite{bonnor1,bonnor2}.  These
Bonnor stars when sufficiently compact show quasiblack hole behavior
\cite{lemosweinberg,kleber,lemoszanchin1,lemoszanchin2008}.  These
stars and quasiblack holes have no matter pressure, only
electromagnetic pressure.  But now, Weyl-Guilfoyle fluids have,
besides electromagnetic pressure, matter pressure. Under the
assumption of spherical symmetry, Guilfoyle \cite{guilfoyle} exhibited
solutions which can be matched to the electrovacuum
Reissner-Nordstr\"om spacetime to yield global asymptotically flat
stars, i.e., charged stars with pressure.

Here we explore one particular class of those spherically symmetric
cold charged fluid stars and show that these stars display quasiblack
hole behavior, i.e., the matter boundary approaches its own
gravitational radius (or horizon) in a well-behaved manner.
Although the cold charged stars have a nonextremal outer metric,
the quasiblack hole regime is extremal always. We
analyze in which cases the energy conditions are obeyed, and also
study a subclass for which the speed of sound in the matter is less
than the speed of light.  So a quasiblack hole with pressure obeying
the necessary physical requirements is presented here. Quasiblack
holes purely supported by electrical charge have been known
\cite{lemosweinberg,kleber,lemoszanchin1} (see also
\cite{bonnor1,bonnor2} with hindsight, and \cite{lemoszanchin2008} for
further results and references). The presence of pressure in
quasiblack hole solutions is important, since it tends to stabilize
the system. Indeed, an important development provided by these
solutions is that without pressure the matter, Majumdar-Papapetrou
matter, is unstable to kinetic perturbations. So the whole solution is
unstable to these perturbations. Charged solutions with pressure, and
along with them quasiblack holes with pressure, avoid this
instability. Thus, if indeed stable, these compact relativistic
charged spheres in the form of cold quasiblack holes with
pressure, are really frozen stars \cite{zeldovichbookorpaper,mtw}. As a
black hole, a quasiblack hole freezes to observers outside, but unlike
a black hole, at the horizon limit, the star is still intact, it does
not collapse.  In the quasiblack hole context, unlike in the black
hole case, the star is not irrelevant. Thus the name frozen star is
appropriate. From a well-motivated chain of works, applying Weyl's
leading method of solving Einstein's equations we thus arrive at the
concept of frozen stars, solutions which exhibit the behavior of
quasiblack holes with pressure. Of course, quasiblack holes considered
as frozen stars are more akin to the dark stars of Michell
and Laplace (see \cite{israel}) than black holes themselves.

The other way to solve Einstein's equations with charged matter with
pressure is through the integration of the TOV equation for the
gradient of the pressure, in conjunction with the other equations for
the other fields.  There are many works that use this method, we only
quote the most relevant to our work. In an important work, de~Felice
and collaborators \cite{felice1,felice2} found relativistic charged
sphere solutions with pressure by the TOV method. Moreover, they took
the limit to the black hole regime, and through numerical integration
still found in this limit a solution. With hindsight this solution is
an extremal quasiblack hole with pressure solution. The equation of
state for the fluid considered is one for an incompressible fluid,
with constant energy density, and in such a case the speed of sound is
infinite, which might be considered a drawback. Bonnor \cite{bonnor2}
took notice of this solution and made comments and comparisons
showing that his previous solution \cite{bonnor1} had indeed the
same gravitational overall properties.  The search of quasiblack hole
solutions through the TOV method is nontrivial. In fact, charged star
solutions with pressure were explored in \cite{malheiro,ghezzi}
without finding the quasiblack hole regime.  There are yet other ways
to solve Einstein's equations. For instance in \cite{lue1,lue2} an
improved method, analytical and numerical, was devised to solve the
Einstein---Yang-Mills--Higgs system of equations. The quasiblack holes
that arose from this system are indeed also extremal quasiblack holes
with pressure since the Yang-Mills--Higgs system has an effective
built-inn pressure on its own. The generic properties of all types of
extremal quasiblack holes with pressure are discussed in \cite{lz5}.

The existence or not of solutions for compact objects in general
relativity is connected with the Buchdahl limit.  Buchdahl
\cite{buchdahl} found for perfect fluids that if the radius $r_0$ and
the mass $m$ of the star is such that $m\geq\frac49\, r_0$, then there
is no equilibrium, see also \cite{karageorgis} for more on this limit.
For instance the Schwarzschild interior solution with constant energy
density for the matter obeys this inequality. If one adds electric
charge to the system the Buchdahl limit is changed, the mass to radius
of the star ratio $\frac{m}{r_0}$ can certainly increase. Indeed the
existence of quasiblack holes shows that the radius of an electrically
charged star can approach its own gravitational radius, $m=r_0$.
Buchdahl limits for charged stars have been worked out in
\cite{yunqiang,mak,giuliani,bohmer} and in particular
\cite{andreasson} gives a sharp limit, $ m\geq \left(\frac{\sqrt
r_0}{3} +\sqrt{ \frac{r_0}{9}+\frac{q^2}{3r_0} }\right)^2$, where $q$
is the total charge of the star. For $q =r_0$, this result admits the
extremal case $m= r_0$ which, as we shall see below, corresponds to
the quasiblack hole limit satisfied by the particular solutions
studied here. Incidentally, the $a\to\infty$ limit of the solutions we
consider yield the Schwarzschild interior solutions.  These uncharged
Schwarzschild interior solutions contain no quasiblack hole, of
course.

Another related issue is concerned with regular black holes.  Regular
black holes are black holes devoid of singularities.  There are two
types of regular black holes. In one type there is a magnetic core
with an event horizon which joins into a magnetically charged vacuum
solution different from the Reissner-Nordstr\"om solution
\cite{bardeen,borde,r4} (see also \cite{bbl} for electrically charged
regular solutions in nonminimal theories).
In the other type, the bulk inside the horizon is
formed of a portion of the de Sitter space which joins into a
Schwarzschild vacuum solution
\cite{dym,galtsovl,lake,r9,bz,zasl}. Curiously, the $a\to\infty$ limit
of the solutions we consider here yields a branch of solutions, other
than the Schwarzschild interior solutions, which are regular
electrically charged black holes.

The present paper is organized as follows. In
Sec.~\ref{sec-basicequations} the basic equations describing a charged
fluid of Weyl-Guilfoyle type are written. A particular spherically
symmetric solution to the equations representing relativistic charged
stars, found by Guilfoyle, is shown in
Sec.~\ref{sec-sphericalsolution}. Then, Sec.~\ref{sec-QBH} is devoted
to the study of the main properties of these stars. We first review
the definition of a quasiblack hole in Sec.~\ref{sec-definition}. Then
we obtain analytically the quasiblack hole limit of the solution in
Sec.~\ref{sec-analytical}. In passing we take the $a\to\infty$ limit
and show that there are two branches of solutions, one corresponds to
the Schwarzschild interior solutions, the other to regular charged
black holes. In Sec.~\ref{sec-numerical} we plot, for three distinct
cases, the relevant curves for each one of the metric and electric
potentials, and for the fluid quantities. All three cases contain
relativistic star solutions and quasiblack hole (frozen star)
solutions. In Sec.~\ref{sec-conclusion} we conclude.

\section{Weyl-Guilfoyle charged fluids: Basic equations}
\label{sec-basicequations}

The cold charged fluids considered in the present work are described by
Einstein-Maxwell equations, which can be written as
\begin{eqnarray}
& &G_\munu=  8\pi \,
\left( T_\munu+ E_\munu\right)\, ,
\label{einst}\\
& & \nabla_\nu F^\munu = 4\pi\,J^\mu\,, \label{maxeqs}
\end{eqnarray}
where Greek indices $\mu, \nu$, etc., run from $0$ to $3$.
$G_\munu=R_\munu-\frac{1}{2}g_\munu R$ is the Einstein tensor, with
$R_\munu$ being the Ricci tensor, $g_\munu$ the metric tensor, and $R$ the
Ricci scalar. We have put both the speed of light $c$ and the
gravitational constant $G$ equal to unity throughout. $E_\munu$ is the
electromagnetic energy-momentum tensor, which can be written in the form
\begin{equation}
4\pi\, E_\munu= {F_\mu}^\rho F_\nu{_\rho} -\frac{1}{4}g_\munu
F_{\rho\sigma} F^{\rho\sigma}\, ,\label{maxemt}
\end{equation}
where the Maxwell tensor is
\begin{equation}
F_\munu = \nabla_\mu {\cal A}_\nu -\nabla_\nu {\cal A}_\mu\, ,
\label{ddemfield}
\end{equation}
$\nabla_\mu$ representing the covariant derivative, and ${\cal A}_\mu$ the
electromagnetic gauge field. In addition,
\begin{equation}
J_\mu = \rho_{\rm e}\, U_\mu\, ,\label{current}
\end{equation}
is the current density, $\rho_{\rm e}$ is the electric charge density,
and $U_\mu$ is the fluid velocity. $T_\munu$ is the material
energy-momentum tensor given by
\begin{equation}
T_\munu = \left(\rho_{\rm m}+p\right)U_\mu U_\nu
+p  g_\munu \, ,\label{fluidemt}
\end{equation}
where $\rho_{\rm m}$ is the fluid matter-energy density, and $p$ is
the fluid pressure.

We assume the spacetime is static and the metric
\begin{equation}
ds^2 = g_{\mu\nu}\,dx^\mu\,dx^\nu\, \label{metrico}
\end{equation}
can be written in the form
\begin{equation}
ds^2 = - W^2 dt^2 + h_{ij}\,dx^idx^j\, ,
\quad\quad i,\,j=1,\, 2,\, 3\,  . \label{metricg}
\end{equation}
The gauge field ${\cal A}_\mu$ and four-velocity $U_\mu$ are then
given by
\beqa
& &{\cal A}_\mu = -\phi\,\delta_\mu^0\, ,\label{gauge1}\\
& &U_\mu =  -W\, \delta_\mu ^0\, .\label{veloc1}
\eeqa
The spatial metric tensor $ h_{ij}$, the metric potential $B$ and the
electrostatic potential $\phi$ are functions of the spatial
coordinates $x^i$ alone.

The particular relativistic
cold charged stars we are going to study here belong to a
special class of systems in which the metric potential $W$ and the
electric potential $\phi$ are functionally related through a
Weyl-Guilfoyle relation (see Guilfoyle \cite{guilfoyle}, see also
Lemos and Zanchin \cite{lemoszanchin2009})
\begin{equation}
W^2= a \left(-\epsilon\,\phi+b\right)^2+c \, ,
\label{weylgrel1}
\end{equation}
with $a$, $b$ and $c$ being arbitrary constants and $\epsilon=\pm 1$,
the parameter $a$ being called the Guilfoyle parameter.
Then, the charged pressure fluid quantities $\rho_{\rm m}$, $p$ and
$\rho_{\rm e}$ satisfy the constraint \cite{lemoszanchin2009}
\beq
a\,b\, \rho_{\rm e}=\epsilon \left[\,\rho_{\rm m}+
3p+\left(1-a\right)\,
\rho_{\rm em}\right]W +\epsilon a\, \rho_{\rm e}\phi \,,
\label{wgcondition1}
\eeq
or
\beq
a\, \rho_{\rm e}\left(-\epsilon\,\phi +b\right)  =
\epsilon\, \left[\rho_{\rm m}+ 3 p+ \epsilon\left(1-a\right)
\rho_{\rm em}\,\right] W \,,
\label{qgcondition2}
\eeq
with $\rho_{\rm em}$ standing for the electromagnetic energy density
defined by
\beq
\rho_{\rm em } = \frac{1}{8\pi}\frac{
\left(\nabla_i\phi\right)^2}{W^2}\,.
\label{emdensity}
\eeq
Such matter systems we call Weyl-Guilfoyle fluids.

\section{Spherical solutions: General analysis}
\label{sec-sphericalsolution}

We repeat in this section one class of the spherically symmetric solutions
found by Guilfoyle \cite{guilfoyle} in order to study its properties.  The
metric (\ref{metrico}) instead of being generically written as
(\ref{metricg}), is more conveniently written in a spherically
symmetric form, namely,
\beq
ds^2 = -B(r)\,dt^2 + A(r)\,dr^2 + r^2\,d\Omega \, . \label{metricsph}
\eeq
where $r$ is the radial coordinate, $A$ and $B$ are functions of $r$
only, and $ d\Omega$ is the metric of the unit sphere $S^{2}$. The gauge
field ${\cal A}_\mu$ and the four-velocity $U_\mu$ are now
\beqa
& &{\cal A}_\mu = -\phi(r)\,\delta_\mu^0\, ,\label{gauge2}\\
& &U_\mu =  -\sqrt{B(r)\,}\, \delta_\mu ^0\, .\label{veloc2}
\eeqa
The cold
charged pressure fluid is bounded by a spherical surface of radius
$r=r_0$, and in the electrovacuum region, for $r>r_0$, the metric and the
electric potentials are given by the Reissner-Nordstr\"om solution
\beqa
B(r)&=&\frac{1}{A(r)}= 1 -\frac{2m}{r}+\frac{q^2}{r^{2}} \,,
\label{f(r)} \label{RNST}\\
\phi(r)& =& \frac{q}{r}+\phi_0\, , \label{phi-RN}
\eeqa
$\phi_0$ being an arbitrary constant which defines the zero of the
electric potential, and that, in asymptotically Reissner-Nordstr\"om
spacetimes as we consider here, can be set to zero.

Now we review the general properties of the spherically symmetric
solutions with the boundary condition given above. The first integral
of the only nonzero component of Maxwell equations \eqref{maxeqs}
furnishes
\beq
Q(r) =  r^{2}\,\frac{\phi^\prime (r)}{\sqrt{B(r) \,A(r)}}\, ,
\label{qspherical}
\eeq
where the prime denotes the derivative with respect to the radial
coordinate $r$ and an integration constant was set to zero. $Q(r)$ is
the total electric charge inside the surface of radius $r$.  The class
of solutions we are interested in here has $c=0$ in Eq.~(\ref{weylgrel1}),
and is classified as class Ia in \cite{guilfoyle}, and so we can write
$\phi (r)$ in terms of $B(r)$ as,
\beq
\epsilon \phi(r) = {b} -\sqrt{\dfrac{B(r)}{a}} \, .
\label{phiofB}
\eeq
With this result, the amount of electric charge
inside a spherical surface of radius $r$, cf. Eq.
\eqref{qspherical}, is
now given by
\beq
Q(r) = \frac{-\epsilon\,r^{2}\,B^\prime(r)} {\,2\sqrt{a A(r)}\, B(r)}\, .
\label{qspherical2}
\eeq

The continuity of the metric functions on the surface $r=r_0$ yields
$B(r_0)= 1/A(r_0)=  1 -\dfrac{2m}{r_0}+ \dfrac{q^2}{r_0^{2}}$. Using
such a boundary condition and Eq.~\eqref{qspherical2} we get
\beqa
\frac{q}{m}&=& \sqrt{\frac{r_0} {2(a-1)m}} \Biggl[ 2(a-1)
-\frac{r_0}{m}+ \Bigr.        \nonumber\\& &
\Bigl.  + \sqrt{4a(a-1)\left(1-\frac{r_0}{m}\right) +
a^2\frac{r_0^2}{m^2}}\, \Biggr]^{1/2},
\label{guilfmass0}
\eeqa
or inverting,
\beq
\frac{m}{q} =   \left(1-a\right) \,\frac{q}{r_0} +
\sqrt{a\left[1+\left(a-1\right)\frac{q^2}{r_0^2}\right]}\,,
\label{guilfmass}
\eeq
where the fact that $Q(r=r_0)= q$ was taken into account.  The
equivalent constraints~\eqref{guilfmass0} or~\eqref{guilfmass} hold
for all spherically symmetric charged pressure fluid distributions
whose boundaries are given by a spherical surface of radius $r=r_0$,
and whose potentials are related as in Eq.~\eqref{phiofB}.  The
extremal relation $q=m$ holds when $a=1$, in which case the relation
among $B$ and $\phi$ is always a perfect square and the $r_0=m$ limit
represents a quasiblack hole.  (In four dimensions, these
Majumdar-Papapetrou charged fluids were studied in
\cite{majumdar,papapetrou}, and their corresponding Bonnor stars and
quasiblack holes in
\cite{bonnor1,bonnor2,lemosweinberg,kleber,lemoszanchin1} as well as
in \cite{lz1,lz2,lz3,lz4}, and in $d>4$ dimensions these fluids were
studied in \cite{lemoszanchin2005} and their corresponding Bonnor
stars quasiblack holes in \cite{lemoszanchin2008}.)  Now, for
$a\neq1$, we are still considering $B=B(\phi)$ as a perfect square
(since $c=0$, see Eq.~(\ref{weylgrel1})) but, unlike a Bonnor star,
the relation $q=m$ does not hold in general. However, from
Eq.~(\ref{guilfmass0}), or Eq.~(\ref{guilfmass}), one sees that, for
$a\neq1$, there is also the possibility in which the equality $m\,
r_0=q^2 $ holds, which may yield back, eventually, the relation
$q=m$. As we will see below, this is the quasiblack hole limit. Since
the inequality $a\neq 1$ holds in this case, it means the
corresponding quasiblack hole is made of a fluid with nonzero
pressure.

Guilfoyle's solutions are found under the assumptions that the
effective energy density $\rho_{\rm eff}(r)=\rho_{\rm m}(r) + \rho_{\rm
em}(r) = \rho_{\rm m}(r) +
\dfrac{Q^2(r)}{8\pi\,r^4}$ is a constant, and that the metric
potential $A(r)$ is a particularly simple function of the radial
coordinate. Namely \cite{guilfoyle},
\beqa
& & 8\pi\,\rho_{\rm m}(r) + \dfrac{Q^2(r)}{r^4} = \frac{3}{R^2}\, ,
  \label{rhoconst} \\
& & A(r) =\left({1 - \dfrac{r^2}{R^2}}\right)^{-1}\, , \label{A(r)1}
\eeqa
where $R$ is a constant to be determined by the junction conditions of
the metric at the surface $r=r_0$. In fact, by joining the function
$A(r)$ in Eq.~\eqref{A(r)1} with the $g_{rr}$ coefficient of the
exterior metric \eqref{RNST}, the constant
$R$ in Eq.~\eqref{rhoconst} is found
in terms of the parameters $r_0$, $m$, and $q$,
through the equation
\beq
\frac{1}{R^2} = \frac{1}{r_0^3}\left( 2m - \frac{q^2}{r_0}\right)\, .
\label{Rm2}
\eeq
With these simplifying  hypotheses, Guilfoyle \cite{guilfoyle} was able
to write some interesting exact solutions.  We shall show that the
class Ia of such solutions admits an extremal case such that $q=m=r_0$,
which represents a quasiblack hole. Let us then write here
this Ia class of solutions,
\beqa
 B(r) & = &\left[\frac{2-a}{a^{1+1/a}}\,F(r)\right]^{2a/(a-2)} \, ,
 \label{B-sol1}  \\
 8\pi \rho_{\rm m}(r)& =& \frac{3}{R^2} -\frac{a}{\left(2-a\right)^2}
\frac{k_0^2\,r^2} {F^2(r)}\, , \label{rhom-sol1} \\
Q(r) &=& \frac{\epsilon \sqrt{a\,}}{2-a}\frac{ k_0\,r^3}{F(r)}
 , \label{charge-sol1}\\
 8\pi p(r) &=& -\frac{1}{R^2} + \frac{a}{\left(2-a\right)^2}
\frac{k_0^2\,r^2} {F^2(r)}      +\nonumber\\ &&
+ \frac{2k_0\,a}{2-a}\frac{\sqrt{1-r^2/R^2}}{F(r)}\, ,
\label{p-sol1}
\eeqa
where $k_0$ is an integration constant, and $F(r)$ and $Q(r)$ are defined
respectively by
\beqa
F(r) &=&  k_0 R^2\sqrt{1 - \frac{r^2}{R^2}}-k_1 \, ,\\
Q(r) & =& 4\pi\int_0^r \rho_{\rm e}(r)
\frac{r^2 dr}{\sqrt{1-\frac{r^2}{R^2}}}= \nonumber\\ &=&
 \frac{r^2}{\sqrt{B(r)}}
\sqrt{1-\frac{r^2}{R^2}} \,
\ \frac{d\phi(r)}{dr} \, ,
\label{charge-def1}
\eeqa
with $k_1$ being another integration constant.
The integration constants $k_0$ and $k_1$ are determined by using the
continuity of the metric potentials $A(r)$ and $B(r)$ and the first
derivative of $B(r)$ with respect to $r$ at the boundary $r=r_0$.
The result is
\beqa
 k_0 &=&\!\!\frac{|q|\,a^{2/a}}{r_0^3}
\left(\frac{m}{q}-\frac{q}{r_0}\right)^{1-2/a}\, ,\label{constk}\\
k_1&=& \!\!\sqrt{1-\dfrac{r_0^2}{R^2}}\!
\left[k_0\, R^2\!- \frac{a^{1+1/a}}{2-a}\!\left(1-\dfrac{r_0^2}{R^2}
\right)^{\!\!-1/a}\right]. \label{constk_1}
\eeqa
From Eqs.~\eqref{charge-sol1} and \eqref{charge-def1} we get both the
electric charge density $\rho_{\rm e}$  and the electromagnetic energy
density $\rho_{\rm em}$,
\beqa
&&\hspace*{-.7cm}8\pi  \rho_{\rm e}(r)  =
\frac{\epsilon\,\sqrt{a\,}}{2-a}
\,\frac{k_0^2\, r^2}{F^2(r)} \left(\frac{\,3F(r)}{k_0}\sqrt{1-
\frac{r^2}{R^2\,}} - 1\right) , \label{rhoe-sol1} \\
&&\hspace*{-.7cm} 8\pi \rho_{\rm em}(r)
=\frac{a}{(2-a)^2}\,\frac{k_0^2\,r^2}{F^2(r)}\, . \label{rhoem-sol1}
\eeqa
Since $c=0$ in the solutions we are considering (see
Eq.~(\ref{weylgrel1})), Eq.~\eqref{wgcondition1} with Eq.~\eqref{weylgrel1}
turns into
\beq
\sqrt{a\,}\,\rho_{\rm e} =\epsilon\left[\rho_{\rm m}+ 3p
+\left(1-a\right)\rho_{\rm em} \right] \, , \label{wgcondition_sol1}
\eeq
which can be interpreted as an equation of state which, as one can
check, is obeyed by the charged fluid represented by the solution presented
above. Furthermore, the above solution is valid for all $a>0$, the case
$a=\infty$ yielding the uncharged ($q=0$) Schwarzschild interior solution.

Another important quantity to determine is the speed of sound within the
fluid. We take the usual definition for the speed of sound $c_{\rm s}$,
\beq
c_{\rm s}^2 = \frac{\delta p}{\delta \rho_{\rm m}}, \label{vsound-def}
\eeq
and consider variations of the pressure $p$ and of the energy density
$\rho_{\rm m}$ in terms of the radial coordinate $r$, i.e., $\delta p
= p^\prime(r) \delta r$, etc., where the prime denotes derivative with
respect to $r$. Hence, we may write $c_{\rm s}^2=
\dfrac{p^\prime}{\rho_{\rm m}^\prime}$.  The result is the speed of
sound as a function of the radial coordinate,
\beq \label{vsound}
c_{\rm s}^2 =   \left|2-a\right|\,
\frac{\,k_2^2- k_2\,\sqrt{1-{r^2}/{R^2}}}{1-k_2\sqrt{1-{r^2}/{R^2}}}-1 ,
\eeq
where we have defined $k_2 = \dfrac{k_1}{k_0R^2}$.

In the following we analyze in some detail the physical properties of
this charged solution, exploiting, in particular, the dependence of the
metric and electric potentials, and of the fluid quantities on the free
Guilfoyle parameter $a$.

\section{Relativistic charged stars and quasiblack holes
with pressure (or frozen stars) }
\label{sec-QBH}

\subsection{Definition of generic relativistic charged stars and
definition of quasiblack holes with pressure (or frozen stars)}
\label{sec-definition}

A generic relativistic cold charged star, or sphere, is here defined
as a smooth ball in which the gravitational, electromagnetic and
matter fields have nonsingular behavior throughout the matter, which
in turn is matched smoothly to the Reissner-Nordstr\"om exterior
solution.

A quasiblack hole or frozen star can also be properly defined. For
such a task we follow Ref.~\cite{lz1} and consider the
metric in the form \eqref{metricsph} with solution given in
Eqs.~\eqref{A(r)1} and \eqref{B-sol1} for the interior which matches
continuously to the exterior metric in the form \eqref{RNST}. Then,
consider the following properties: (a) the function $1/A(r)$ attains a
minimum at some $r^{\ast }\neq 0$, such that $1/A(r^{\ast
})=\varepsilon $, with $\varepsilon <<1$, this minimum being achieved
either from both sides of $r^{\ast }$ or from $r>r^{\ast }$ alone, (b)
for such a small but nonzero $\varepsilon $ the configuration is
regular everywhere with a nonvanishing metric potential $B(r)$, and (c)
in the limit $\varepsilon \rightarrow 0$ the potential $B(r)
\rightarrow 0$ for all $r\leq r^{\ast }$. These three features define
a quasiblack hole and, in turn, entail some nontrivial
consequences:
(i) there are 3-volume regions, rather than 2-surface regions (as in
the black hole case), for which the redshift is infinite;
(ii) when $\varepsilon
\rightarrow 0$, a free-falling observer finds in his own frame
infinitely large tidal forces in the whole inner region, although the
spacetime curvature invariants remain perfectly regular everywhere;
(iii) in the limit, outer and inner regions become mutually
impenetrable and disjoint; and one can also show that (iv) for external
far away observers the spacetime is virtually indistinguishable from
that of an extremal black hole.  In addition, well-behaved
(with no infinite surface stresses) quasiblack holes must be
extremal. The quasiblack hole is on the verge of forming an event
horizon, but it never forms one, instead, a quasihorizon appears. For
a quasiblack hole the metric is well-defined and everywhere regular.
Nonetheless, the properties that arise when $\varepsilon=0$ have to be
examined with care for each model in question.

\subsection{Analytical study of quasiblack holes
with pressure (or frozen stars)}
\label{sec-analytical}

An interesting property of the solution presented in
Sec.~\ref{sec-sphericalsolution}
(cf. Eqs.~\eqref{A(r)1}-\eqref{p-sol1}) is that all the physical
quantities are well-behaved even in the quasiblack hole limit. In what
follows we first verify that the quasiblack hole limit can be
attained, and then we check that in such a limit the physical
quantities related to the charged fluid are well-defined.

\subsubsection{The metric potentials and the electric potential}
\label{sec-coeffs}

Let us show here that the quasiblack hole really exists as a special
limit of the relativistic charged star solution presented
above. According to the definition of the preceding section, the
quasiblack hole should be extremal, so that the mass $m$ approaches
the charge from above, $m\rightarrow q^+$. In such a limit, the
exterior metric $\eqref{RNST}$ tends to the quasi-extremal
Reissner-Nordstr\"om metric, for which the two horizons $r_\pm = m\pm
\sqrt{m^2 - q^2}$ are very close to each other, $r_+\sim m \sim r_-$.
Moreover, there must be a quasihorizon $r^\ast$, and then the radius
of the star $r_0$ must coincide with $r^\ast$. Hence, the quasiblack
hole limit of the star corresponds to the limit $r_0 \rightarrow r_+$
from above.  We can then assume a relation of the form $q\sim \left(1
- \sqrt\varepsilon\right)r_0$, for a small nonnegative
$\varepsilon$. This means that the quasiblack hole limit corresponds
to the most compact charged star we can have, with $r_0/m$ close to
unity. In the present case, the mass $m$ and the charge $q$ are
related by Eq.~\eqref{guilfmass}, which implies in $m/q\sim 1
+(a-1)\varepsilon/2a$, and then $m/r_0 \sim 1 -\sqrt\varepsilon$. From
these results we may obtain the quasiblack hole limit for all other
quantities. For instance, the difference $m/q-q/r_0$ which appears in
the constants $k_0$, $k_1$, etc., is of the order of
$\sqrt\varepsilon$. Taking these considerations into the corresponding
equations we find, for instance, $R^{-2}\sim (1-
2\sqrt\varepsilon)/r_0^2 $, $k_0 \sim
a^{2/a}\varepsilon^{(a-2)/2a}/r_0^2$, $B(r_0) \sim \varepsilon$,
$k_1\sim a^{(a+4)/2a}\varepsilon^{(a-2)/2a}\,/(2-a)$. The function
$F(r)$ goes as $k_1$ for all $r$ inside the star, and then
Eq.~\eqref{B-sol1} implies that $B(r)$ is of the order of
$\varepsilon$ for all $r$ in the interval $0\leq r \leq
r_0$. Moreover, we find $1/A(r_0)\sim\varepsilon$, satisfying the
properties of a quasiblack hole as defined in
Sec.~\ref{sec-definition}.  In the quasiblack hole limit, the metric
potentials become
\beqa
\hspace*{-.3cm}&B(r) &= \left\{\begin{array}{l}
\left(1 +\dfrac{2-a}{\sqrt{a\,}}
\sqrt{1-\dfrac{r^2}{r_0^2}}\right)^{\frac{2a}{a-2}}\!\!\!
\dfrac{\varepsilon}{a}\,,\;\;
r\leq    r_0\, ,\\
\left(1 - \dfrac{r_0}{r}
\left[1-\dfrac{\sqrt\varepsilon}{\sqrt{a\,}}\right]\right)^2,\,
\quad r\geq r_0\, ,
\end{array} \right. \label{B(r)-qbh} \\
\hspace*{-.3cm}&A(r)^{-1} & = \left\{\begin{array}{c}
\left(1-\dfrac{r^2}{r_0^2}\left[1-\dfrac{\varepsilon}{a}
\right]\right)\, ,
\quad r\leq r_0\, ,\\
\left(1 - \dfrac{r_0}{r}
\left[1-\dfrac{\sqrt\varepsilon}{\sqrt{a\,}}\right]\right)^{2},\, \quad
r\geq
r_0\, ,
\end{array}\label{A(r)-qbh} \right.
\eeqa
where we have used the fact that $ |q| \simeq m = (1 -\sqrt
\varepsilon)\,r_0$ and have written $|q|/r_0^3 = 1/r_0^2$.  The electric
potential $\phi(r)$ at the quasiblack hole limit is obtained from the
above result for $B(r)$ and from Eq.~\eqref{phiofB}.

\subsubsection{The fluid quantities}
\label{sec-fluidqts}

The study of the physical properties satisfied by the energy density
and by the pressure for the present solution was in part done in Ref.
\cite{guilfoyle}. There are, however, further interesting aspects that
should be considered.

We can for instance obtain the analytical expressions at the
quasiblack hole limit for the fluid quantities like mass density,
charge, and pressure,
\beqa
8\pi\rho_{\rm m}(r)\!&\! =\!&\! \dfrac{3|q|}{r_0^3}
-  \dfrac{q^2 r^2}{r_0^6\,} \left(1 +\frac{2-a}{\sqrt{a\,}}\,
\sqrt{1-\dfrac{r^2}{r_0^2}}\right)^{\!-2} ,
\label{rhom-qbh} \\
Q(r) &\!=\!&  \dfrac{q\, r^3}{r_0^3} \left(1
+\frac{2-a}{\sqrt{a\,}}\, \sqrt{1-\dfrac{r^2}{r_0^2}}\right)^{\!-1} ,
\label{charge-qbh}\\
 8\pi p(r) &\!=\!& -\frac{|q|}{r_0^3} + \frac{q^2 r^2}{r_0^6\,}
\left(1 +\frac{2-a}{\sqrt{a\,}}\,
\sqrt{1-\dfrac{r^2}{r_0^2}}\right)^{\!-2}  +\nonumber\\ &&
+ \dfrac{2\,a\, |q| }{r_0^3}\,\left({2-a}
+\sqrt{a\,}/\sqrt{1-\dfrac{r^2}{r_0^2}}\right)^{\!-1} .
\label{p-qbh}
\eeqa
The quasiblack hole limit of the other fluid quantities, namely, of
the charged density $\rho_{\rm e}$ and of the electromagnetic density
$\rho_{\rm em}$ are easily obtained, respectively, from
Eq.~\eqref{wgcondition_sol1} and from the identity $\rho_{\rm em}=
Q^2(r)/8\pi r^4$, and then we do not write them here.

Let us now analyze the speed of sound $c_{\rm s}$ within these
relativistic charged stars in the quasiblack hole limit. For that we
use Eq.~\eqref{vsound} and get the limiting values of the constants
$R^2$, $k_0$, and $k_1$, as done in Sec.~\ref{sec-coeffs}. Then we find
\beq
\label{vsoundQBH}
c_{\rm s}^2(r) =   \frac{{a\,}+ (2-a)\sqrt{a\,}\sqrt{1-{r^2}/{r_0^2}}}
{2-a +\sqrt{a\,} \sqrt{1-{r^2}/{r_0^2}}}-1 .
\eeq
As expected the speed of sound of the quasiblack hole is zero if
$a=1$. At the surface, $c_{\rm s}^2$ tends to $-1 +a/(2-a)$ which
reaches unity for $a=4/3$.  The function $c_{\rm s}^2(r) $ at the center
($r=0$) of the quasiblack hole is such that $c_{\rm s}^2(0)$ is
bounded to $-1$ from above as $a$ tends to zero. In fact we see that
$c_{\rm s}^2(0)$ is zero for $a=1$, and tends monotonically to $-1$ as
$a$ goes to zero. Hence, $c_{\rm s}$ is undefined for all $a<1$. One
can further impose that the speed of sound is smaller than the speed
of light to yield a further interesting class of solutions.\\

\subsubsection{The mass to radius relation and
the $a\to\infty$ limit (the Schwarzschild interior and
the regular black hole branches)}

The mass $m$, the charge $q$, and the radius $r_0$ of the relativistic
stars analyzed here are related by Eq.~\eqref{guilfmass0}, or
equivalently, by Eq.~\eqref{guilfmass}. Hence, besides the parameter
$a$, out of the three quantities $m$, $q$, and $r_0$, we are left with
two more free parameters. One can consider various possibilities. One
possibility is to normalize the quantities in terms of the mass $m$.
This choice is interesting, since the uncharged limit (the
Schwarzschild interior solution) is easily obtained by taking the
limit $a\to\infty$. Therefore, in order to observe this limit to the
(uncharged) Schwarzschild stars, we will in general normalize the
quantities in terms of the mass $m$ of the star. Another possibility
is to normalize the quantities in terms of the charge $q$. This choice
is also interesting since from it one can read directly the star's
mass to radius relation, $m\times r_0$.  Of course, other possibilities
could be considered.

In order to see the difference between the two normalizations just
mentioned we plot Fig.~\ref{mttor0}.  The plot on the left-hand side
of Fig.~\ref{mttor0} shows the behavior of $q/m$ as a function of
$r_0/m$. From the curves, obtained from Eq.~\eqref{guilfmass0}, we
read that, for a fixed mass, the electric charge of the star decreases
with its radius, the maximum value being $q/m=1$ for $r_0/m=1$, the
quasiblack hole limit.  The plot on the right-hand side of
Fig.~\ref{mttor0} gives the mass to charge ratio $m/q$ for a few
values of the parameter $a$ as a function of the normalized radius of
the star $r_0/q$, whose curves are obtained from
Eq.~\eqref{guilfmass}. It shows that, for a fixed charge, the radius
of the star decreases as the mass decreases. This mass to radius
relation behavior is analogous to the behavior of the mass to radius
relation in main sequence stars, and contrary to the mass to radius
relation in white dwarfs.

Another point to be noted is that these kind of stars bear a
relatively large charge to mass ratio. The ratio $m/q$ runs from
unity, the minimum value, at the quasiblack hole limit, to $\sqrt{a}$
for an extremely sparse star, with $r_0 \rightarrow\infty$. In
fact, we see from Eq.~\eqref{guilfmass} that, for fixed $a$, the
maximum value of $m/q$ is found in the limit $r_0\rightarrow\infty$,
and it is given by $ \displaystyle{
 \left.\frac{m}{q}\right|_{\rm max}=\sqrt{a\,}}$ .

Now we show that in the limit of $a\to\infty$ one recovers the
Schwarzschild interior solution $q=0$ and $\rho_{\rm m}={\rm const}$,
and also picks up another branch.  Indeed, taking $a\to\infty$ in
Eq.~\eqref{guilfmass0} we find
\beq
\displaystyle{\lim_{a\to\infty} \dfrac{q^2}{m^2}= \frac{1}{2}\left(2
+\left|\frac{r_0}{m} -2\right| - \frac{r_0}{m} \right)\frac{r_0}{m}}\, .
\label{limsch}
\eeq
Because of the absolute value $\left|{r_0}/{m} -2\right|$ one concludes,
remarkably, there are two branches, ${r_0}/{m}>2$ and
${r_0}/{m} < 2$.

For  ${r_0}/{m}>2$ one finds
\beq
{\lim_{a\to\infty} \dfrac{q^2}{m^2}= 0},
\label{limschindeed}
\eeq
which gives the uncharged Schwarzschild interior solution.

For  ${r_0}/{m}<2$ one finds
\beq
{\lim_{a\to\infty} \dfrac{q^2}{m^2}=
\left(2 -\frac{r_0}{m}\right) \frac{r_0}{m}}\,.
\label{limschother}
\eeq
This solution is weird in this context.
Equation \eqref{limschother} is equivalent to
\beq
1 -\frac{2m}{r_0}+\frac{q^2}{r_0^{2}}=0\,,
\label{horizon}
\eeq
which shows that, for this branch, when $a\to\infty$ for given $m$ and
$q$ there are charged solutions in which the radius of the star $r_0$
is the horizon radius. For $1< r_0/m<2$ these are nonextremal
solutions which represent nonextremal regular black holes with tension
matter.  For $r_0/m=1$ these are extremal solutions which represent
extremal regular black holes with tension matter. Note that these
extremal solutions can be obtained from Eq.~(\ref{B(r)-qbh}) when one
takes the limit $a\to\infty$ and absorbs the $\varepsilon$ term into
the time coordinate so that $B(r)$ is continuous at $r_0$.  This
$a\to\infty$ branch, representing regular black holes rather
than quasiblack holes, should be carefully handled and we do not
explore it further here. Regular black holes either with a charged
core \cite{bardeen,borde,r4,bbl} or with a de Sitter core
\cite{dym,galtsovl,lake,r9,bz,zasl} are known, but with charge and a
de Sitter core together, as found here, seem to have not been
explored.

\begin{widetext}
\vspace*{-.3cm}
\begin{figure}[htb]
\begin{center}
\includegraphics[width=7.2cm,height=4.7cm]{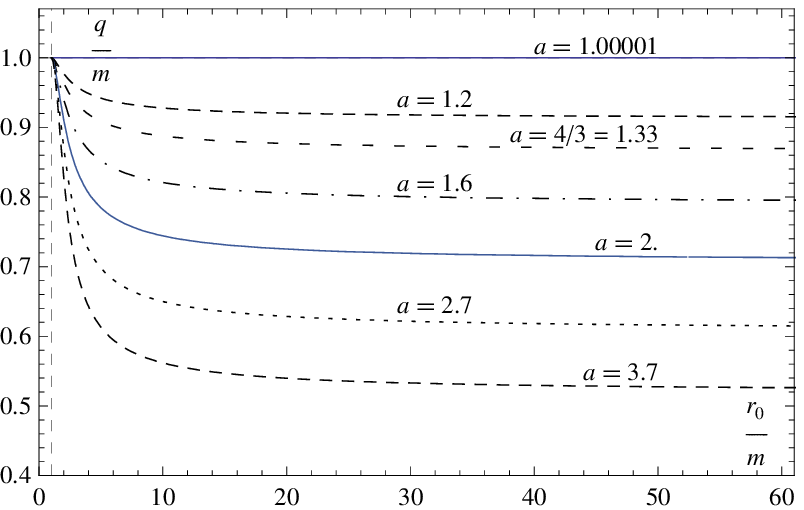}\hskip1.5cm 
\includegraphics[width=7.2cm,height=4.7cm]{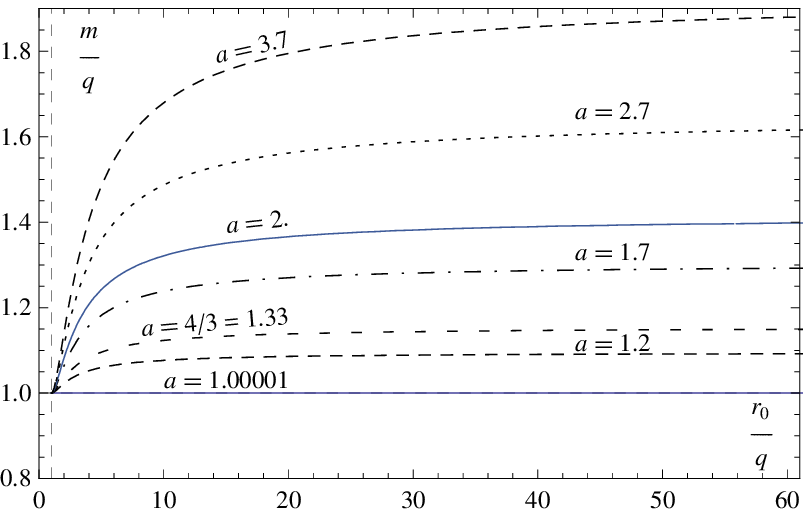}
\caption{
Left: the charge of the star (normalized to the mass) as a function of
the radius of the star (normalized to the mass) for a few values
of $a$ in the interval of interest, $1\leq a\leq 4$.
Right: the mass of the star (normalized to the charge) as a function
of the radius of the star (normalized to the charge) for a few values
of $a$ in the interval of interest, $1\leq a\leq 4$. }
\label{mttor0}
\end{center}
\end{figure}
\end{widetext}

\subsection{Numerical study of relativistic charged
stars and quasiblack holes
(or frozen stars): Three typical cases}
\label{sec-numerical}

\subsubsection{Intervals of $a$}
\label{sounds}

Although valid for all $a>0$, there are some special intervals in the
domain of the parameter $a$ for which the solutions can be considered
more physical. It is known that the Schwarzschild interior solution,
$a=\infty$, having an incompressible fluid as matter source, yields
violation of the dominant energy condition as well as an infinite
speed of sound $c_{\rm s}$, i.e., a speed greater than the speed of
light, bringing into question causality issues. Thus, when discussing
the interesting intervals of the Guilfoyle parameter $a$, which yield what
might be considered physical solutions, it is important to take into
account the energy conditions and the behavior of the speed of sound.
It is a straightforward task verifying that, within a
certain range of parameters, the fluid quantities satisfy the energy
conditions for the whole star, even in the quasiblack hole limit.
Let us then find such a interval.

First we note that the most restrictive conditions arise in the
quasiblack hole limit, and then we study the energy conditions for
this kind of frozen stars. In a charged static fluid, besides the
fluid energy density and pressure, there are the electromagnetic
energy density $\rho_{\rm em}$, the radial electric pressure
$-\rho_{\rm em}$ and the tangential electromagnetic pressures
$\rho_{\rm em}$. It is then useful to define the effective energy
density and pressures of the charged fluid by
\beqa
\rho_{\rm eff}(r)& \equiv&  \rho_{\rm m}(r)
+\rho_{\rm em}(r)=  \rho_{m}(r) + \frac{Q(r) ^2}{8\pi r^4}\, ,\\
p^r_{\rm eff}(r) &\equiv&  p(r) - \rho_{\rm em}(r) = p(r) -
\frac{Q(r) ^2}{8\pi r^4} \, ,\\
p^t_{\rm eff}(r) &\equiv&  p(r) + \rho_{\rm em}(r) = p(r) +
\frac{Q(r)
^2}{8\pi r^4} \, ,
\eeqa
where the upperscript ``$r$'' and ``$t$'' in the above definitions
stand for radial and tangential pressures, respectively.  Therefore,
testing the energy conditions in the present case leads us to check
inequalities such as
\begin{itemize}
\item[(a)]  $\rho_{\rm eff}(r) \geq 0 $,
\item[(b)] $ \rho_{\rm eff}(r) + p^r_{\rm eff} (r) \geq 0 $, or
equivalently, $\rho_{\rm m}(r) + p (r) \geq 0 $,
\item[(c)] $\rho_{\rm eff }(r) + p^t_{\rm eff}(r) \geq 0$,
\item[(d)] $\rho_{\rm eff}(r) + p^r_{\rm eff}(r) + 2p^t_{\rm eff}(r)
\geq 0$,
\item [(e)] $ \rho_{\rm eff}(r) \geq |p^r_{\rm eff}(r)|$,
\item [(f)] $ \rho_{\rm eff}(r) \geq |p^t_{\rm eff}(r)| $, or
equivalently, $ \rho_{\rm m}(r) \geq |p(r)| $.
\end{itemize}
These are requirements for the weak
energy condition (WEC) [(a),(b), and (c)], null energy condition
(NEC) [(b) and (c)], strong energy condition (SEC) [(b), (c), and
(d)], and dominant energy condition (DEC) [(e) and (f)].

Condition (a) is promptly verified after Eq.~\eqref{rhoconst}.
Condition (b) in the form $\rho_{\rm m}(r) + p (r) \geq 0$ implies the
charged fluid satisfies the inequality conditions (c) and (d).  In
addition, when the charged fluid satisfies the second inequality of (f)
it also satisfies (e).  Therefore, it is convenient to check first the
conditions on the matter quantities $\rho_{\rm m}(r)$ and $p(r)$ and
some relations between them.

Starting with the matter-energy density $\rho_{\rm m}(r)$, a numerical
analysis shows that in the case $r_0/m=1.00001$ (the quasiblack hole limit)
one has $\rho_{\rm m}(r)> 0$ for all $r$ if $a$ is in the interval $0<
a\lesssim 3.53523$. For $ 3.53523\lesssim a \leq 4$, $\rho_{\rm m}(r)$
is negative in some intervals of $r$ inside the star, the
corresponding (negative) minimum of $\rho_{\rm m} (r)$ moves toward
the center of the star while $a$ increases. For $a=4$, the minimum
value of $\rho_{m}(r)$ shifts to $r\simeq 0.017$ and becomes very
large (negative) but finite. For $a>4$, at the quasiblack hole limit
$\rho_{\rm m}(r)$ behaves wildly with $r$. The extremal negative value
of $\rho_{\rm m}(r)$ becomes arbitrarily large, moves toward the
surface $r=r_0$ as $a$ grows, and eventually disappears for very large
$a$.

Now we turn to the behavior of the matter pressure $p(r)$ at the
quasiblack hole limit $r_0/m=1.00001$. It is negative for all $a$
belonging to the interval $0< a<1$, vanishing for $a=1$ in which case
the matter is composed of charged dust. The pressure $p(r)$ is
positive in the interval $1< a \leq 4$, reaching a very large finite
positive value at $a = 4$. In the interval $4< a <\infty $, the
behavior of the pressure is as wild as the energy density $\rho_{\rm
m}(r)$, the difference being that $p(r)$ reaches arbitrarily large
positive values at points where the energy density gets arbitrarily
large negative values.

It is also useful to compare $\rho_{\rm m}(r)$ with $p(r)$.  For small
$a$ the absolute value of $p(r)$ is smaller than $\rho_{\rm m}(r)$. In
particular, the central pressure is smaller than the central energy
density, the equality $p(r=0)=\rho_{\rm m}(r=0)$ being reached at $a=
\left(1+\sqrt{13\,}\right)^2/9 \simeq 2.35679$.  For $a$ larger than
that value, the central pressure is always larger than the central
energy density. A more detailed analysis valid for any $r$, not just
$r=0$, shows that at the quasiblack hole limit $r_0/m=1.00001$, the
equality $p=\rho_{\rm m}$ is reached at $a\simeq2.32665$, above this
value one has $p(r) > \rho_{\rm m}(r)$ at some interval of values of
$r$ inside the quasiblack hole or frozen star.
Another quantity of importance involving $\rho_{\rm m}(r)$ and $p(r)$
together is $\rho_{\rm m}(r) + p(r)$ for which we obtain that it is
positive for all stars and for the frozen star if $a$ is in the interval
$0<a\leq 4$.

Finally, let us stress that the analysis was performed for $r_0/m =
1.00001$, i.e., in the quasiblack hole limit. On the other hand,
sufficiently sparse stars, $r_0/m > 1 $, satisfy all of the energy
conditions for all $a$.

In summary, after a careful numerical analysis of the quasiblack hole
limit, we can state the following:
\begin{itemize}
\item[(i)] For $0 < a \leq 4$
conditions (b) and (c) are satisfied
for all $r_0/m$ including the quasiblack hole. Therefore, the charged
fluid satisfies the NEC as long as the Guilfoyle
parameter $a$ is in the interval $0 < a \leq 4$.
\item[(ii)] The WEC requires that conditions (a), (b)
and (c) are satisfied. Hence it is satisfied by all of the stars and
by the quasiblack hole or frozen star if $0< a \leq 4$.
\item[(iii)] The SEC requires conditions (b), (c)
and (d) to be satisfied, and so the charged fluid satisfies the SEC for $a$
in the interval $0 < a \leq 4$.
\item[(iv)]  Conditions (e) and (f) are required by the DEC.
A numerical analysis yields $\rho_{\rm m}(r) \geq
|p(r)| $ for $ 0< a\lesssim 2.32665$ and, within this interval of
$a$, condition (e) is satisfied too. Hence the DEC
is satisfied in the interval $ 0< a\lesssim 2.33$,
where we put $2.32665\simeq 2.33$ from here onwards.
\end{itemize}

Now, we analyze the speed of sound. From Eq.~(\ref{vsound}) we see that
$c_{\rm s}^2(r)$ is a monotonically increasing function of the radial
coordinate $r$. Hence, even if at the center of the star, the parameters
are fixed so that the speed of sound is smaller than the speed of light, it
may happen that $c_{\rm s}(r)$ reaches values greater than unity at some
point $r$ inside the star. As the numerical analysis has shown, the
restriction that the speed of sound is at most as large as the speed of
light imposes the strongest bounds on the range of values of $a$.  Indeed,
such an imposition restricts the allowed values of the parameter $a$ to be
in the interval $1\leq a\leq 4/3$. This is confirmed by the curves shown
in Fig.~\ref{soundspeedcenterandsurface}, where we plot $c_{\rm s}(r)$ as a
function of $a\geq 1$, at the center and at the surface of the star, for
several different values of the parameter $r_0/m$, which represents the
compactness of the star. The curves terminate at $a=1$ because
$dp(r)/d\rho_{\rm m}(r)$ is negative in the interval $0<a<1$, and so the
speed of sound is not defined in that interval. The value $a=4/3$, shown as
a vertical dashed line in Fig.~\ref{soundspeedcenterandsurface}, is a
critical value in the sense that for a very sparse star, i.e., in the limit
$r_0 \rightarrow \infty$, the speed of sound at the center of the star
reaches unity for $a=a_c =4/3$.  It is also seen from
Fig.~\ref{soundspeedcenterandsurface} that, for each $a$, the speed of
sound at the center of the star strongly depends on the compactness
parameter $\beta$. Interestingly, more compact stars have smaller speeds of
sound at the center.  On the other hand, as seen from the right plots in
Fig.~\ref{soundspeedcenterandsurface}, for $a$ in the interval $1\leq a
\leq 4/3$, the speed of sound close to the surface of the star is
practically the same for all $r_0/m$.

\begin{widetext}
\vspace*{-.4cm}
\begin{figure}[htb]
\begin{center}
\includegraphics[width=7.3cm,height=4.3cm]{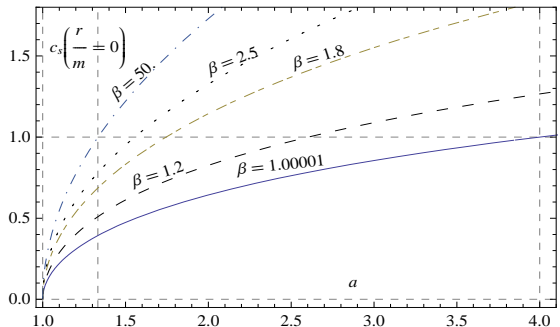}
\hspace*{1.9cm}
\includegraphics[width=7.3cm,height=4.3cm]{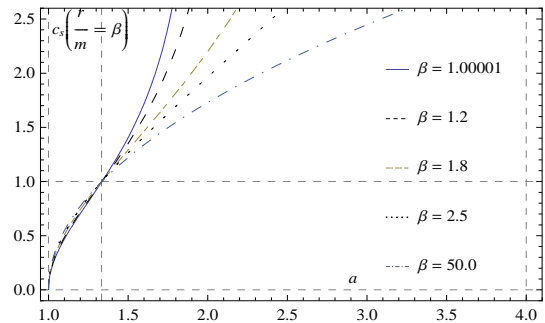}
\caption{On the left, the speed of sound at the center of the
star, $c_{\rm s}(0)$, as a function of the parameter $a\geq 1$, for a few
different values of $\beta = r_0/m$. For $a$ in the interval $0<a<1$,
the speed of sound is not defined, since $dp/ d\rho_{\rm m} <0$. The value
$a_c=4/3=1.33$ is shown as a vertical dashed line. Notice that for a very
sparse star, i.e., in the limit $r_0 \rightarrow \infty$, the speed of
sound at the center of the star reaches unity for $a=a_c =4/3$. For
$a\to\infty$, the Schwarzschild interior solution, $c_{\rm s}(0)$,
is infinite at the center. On the right, the speed of sound
 at the surface of the star, $c_{\rm s}(r_0)$, as a function of the
parameter $a\geq 1$. The value $a_c=4/3$ is shown as a vertical dashed
line. In all the cases, i.e., for all values of $\beta$, the speed of sound
close to the surface of the star tends to unity at $a=a_c$. For
$a\to\infty$, the Schwarzschild interior solution, the speed of sound is
infinite at the surface (in fact the speed of sound in the Schwarzschild
interior solution is infinite throughout the whole fluid region).}
\label{soundspeedcenterandsurface}
\end{center}
\end{figure}

\end{widetext}

For each value of the Guilfoyle parameter $a$ there is an infinity of
star solutions, functions of the compactness of the stars themselves,
i.e., functions of $r_0/m$.  The parameter $a$ is related to the
pressure: for $a<1$ the stars are supported by tension; for $a=1$ the
stars have no pressure, they are Bonnor stars
\cite{bonnor1,bonnor2,lemosweinberg,kleber,lemoszanchin1,lemoszanchin2008};
and for $a>1$ the stars are supported by pressure.  In detail, $a$ can
be divided into the following intervals: $0<a<1$ yields tension stars;
$a=1$ yields charged dust stars; $1<a\leq4/3$ yields stars and
quasiblack holes which obey the energy conditions and have a speed of
sound less than the speed of light; $4/3<a\lesssim2.33$ yields stars
and quasiblack holes which obey the energy conditions and have a speed
of sound greater than the speed of light; $2.33<a\leq 4$  yields
stars and quasiblack holes which obey all the energy conditions but the
DEC and have a speed of sound greater than the speed of light; and
$4\leq a\leq\infty$ yields normal stars which obey all the energy
conditions but the DEC, but yields no quasiblack holes. In this latter
interval of $a$ the matter behaves as in the Schwarzschild solution
($a\rightarrow\infty$), in the sense that before the gravitational
radius is reached gravitational collapse ensues.

The interval $0<a<1$ does not interest us here because it gives
tension stars, the value $a=1$ yields charged dust Bonnor stars
studied previously, and the interval $4< a\leq\infty$ does not yield
quasiblack holes and so again is of no interest here, although very
interesting in other contexts.  The intervals of $a$ that yield
systems that can be pushed into quasiblack holes is $1\leq a\leq
4$. So, since $a=1$ has been studied in previous works, we will study
the interval $1<a\leq4/3$, as well as the intervals
$4/3<a\lesssim2.33$, and $2.33<a\leq 4$.  In order to see the main
features of these charged stars and, in particular, to see the
quasiblack hole limit, we have plotted some important curves for three
typical cases within each interval, namely, $a=1.2$, $a=1.7$, and
$a=3$.  We first verify that the quasiblack hole limit can be
attained, and then we check that in such a limit the physical
quantities related to the charged fluid are well-defined.\\

\subsubsection{The interval
$1< a\leq 4/3$. Typical case: $a=1.2$}
\label{sec-a1.2}

The metric and electromagnetic fields, and the fluid quantities, such as
mass density $\rho_{\rm m}(r)$, charge density $\rho_{\rm e}(r)$, pressure
$p(r)$, and speed of sound $c_{\rm s}(r)$, are well-behaved functions of
the radial coordinate for $1\leq a\leq 4/3$, even in the quasiblack hole
limit. In fact, we have numerically analyzed each one of these functions
for several values of $r_0/m$ as functions of the normalized radial
coordinate $r/m$. Within this interval we analyze the typical case $a=1.2$.

\begin{widetext}

\begin{figure}[ht]
\vskip -.4cm
\begin{center}
\includegraphics[width=6.9cm,height=11.3cm]{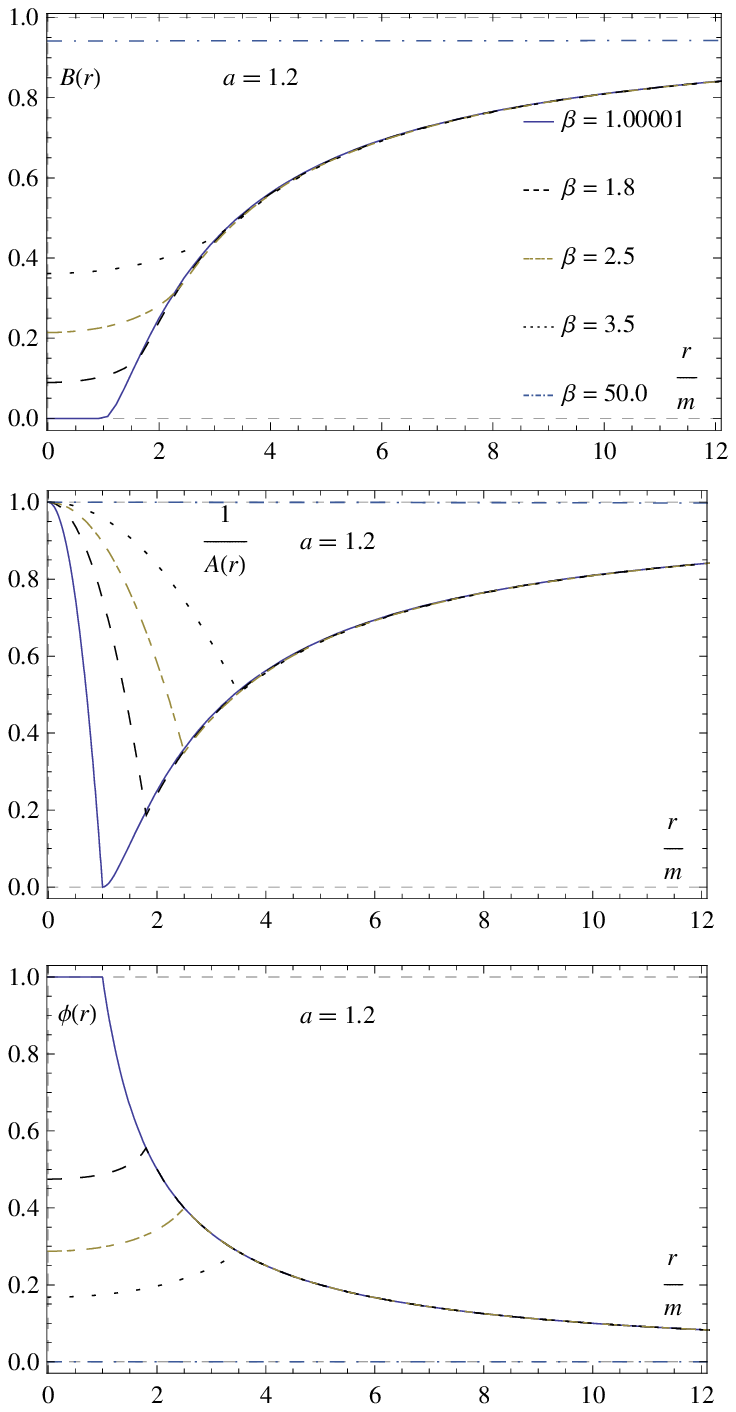}\hskip 1.3cm
\vspace*{-.0cm}
 \includegraphics[width=6.9cm,height=11.3cm]{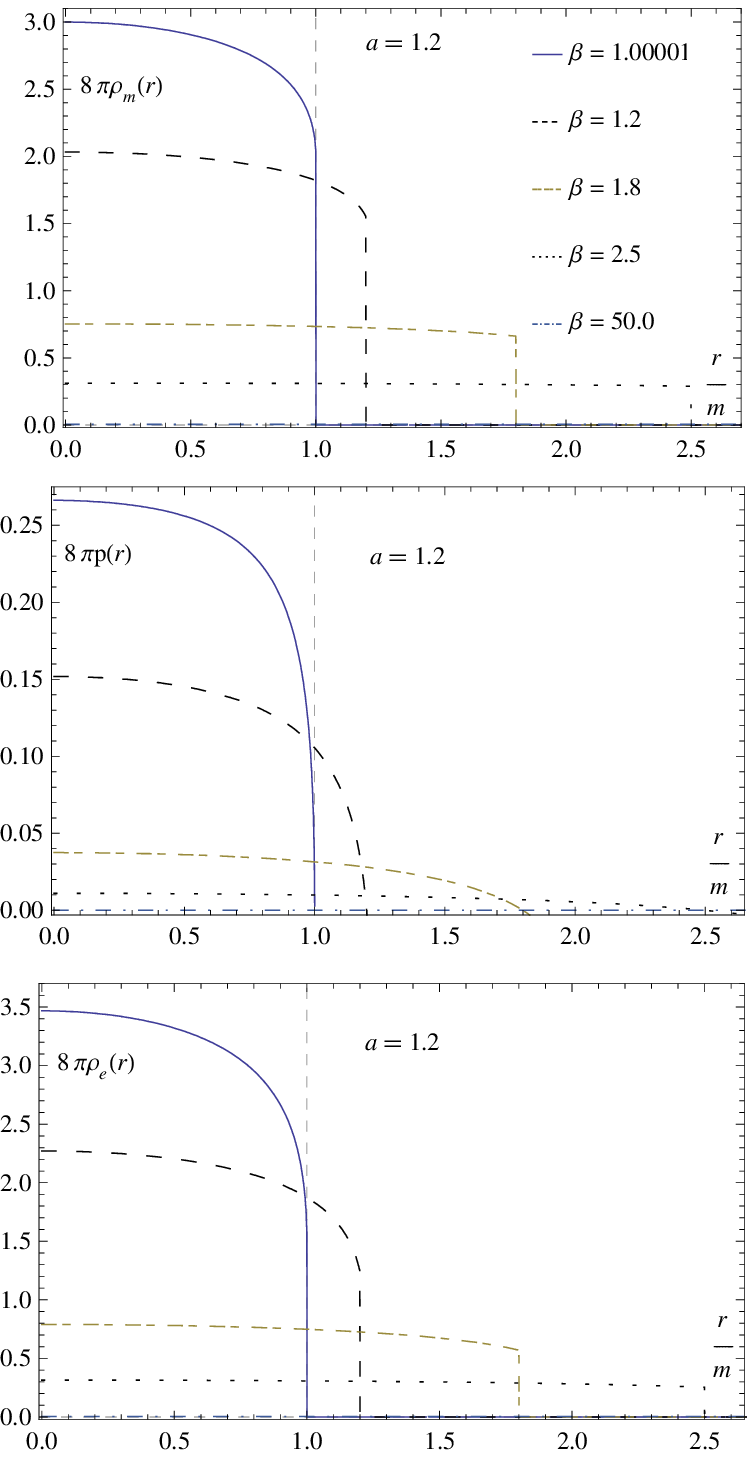}
\vskip -.3cm
\caption{The potentials and fluid quantities for the case $a=1.2$. The
metric potentials $B(r)$ and $1/A(r)$, and the electric potential $\phi(r)$
(left panel) and the fluid quantities $\rho_{\rm m}(r)$, $p(r)$ and
$\rho_{\rm e}(r)$ (right panel). All quantities are plotted in terms of the
normalized radial coordinate $r/m$, where $m$ is the mass of the star,
which is kept fixed, for five values of $\beta= r_0/m$ in each graph:
$\beta=1.00001$ (solid line) $\beta=1.2$ (space-dashed line), $\beta=1.8$
(dashed line), $\beta=2.5$ (dotted line), and $\beta=50.0$ (dot-dashed
line). Notice that the case $\beta=1.2$ is not shown for the potentials,
the case $\beta =3.5$ being shown instead, because the curves of the
potentials for $\beta=1.2$ practically coincide with the curves for $\beta
= 1.00001$. The horizontal straight dotted line represents the asymptotic
limit of the metric potentials for large $r$, $B(r)=1/A(r)=1$, and
$\phi(r)=0$,  and is plotted for comparison. The case $\beta=1$, which
gives $q=m=r_0$, is a quasiblack hole.}
\label{fig-a1pt2}
\end{center}
\end{figure}

\end{widetext}

{\it Metric and the electric potentials}: The metric potentials $B(r)$ and
$1/A(r)$, and the electric potential $\phi(r)$ as a function of the
normalized radial coordinate $r/m$, for the case $a=1.2$, are shown in the
left panel of Fig.~\ref{fig-a1pt2}. The metric potentials $B(r)$ and
$1/A(r)$ are obtained from Eqs.~\eqref{B-sol1} and \eqref{A(r)1},
respectively, while $\phi(r)$ is obtained using relation \eqref{phiofB}
with $c=0$, and by choosing the constant $b$ so that the potential
$\phi(r)$ is a continuous function at the surface $r=r_0$. The exterior
metric is the Reissner-Nordstr\"om metric, and then the curves for the
potential $B(r)$ and $1/A(r)$ tend to unity for large values of the radial
coordinate $r/m$. We plot the curves for each one of the potentials for
five different values of the normalized radius of the star $\beta = r_0/m$,
namely, $\beta= 1.00001$ (solid line), $\beta = 1.8$ (space-dashed line),
$\beta = 2.5$ (dashed line), $\beta = 3.5$ (dotted line), and $\beta =50$
(dash-dotted line). The case $\beta =50$ represents a very sparse star and
the metric potentials are nearly constant close to unity in the hole region
inside the star. As a consequence, the curves of the potentials in this
case are horizontal lines, with $\phi(r)$ very close to zero. On the
other hand, the electric potential is close to zero inside the star. It is
also seen that for the quasiextremal case where $r_0/m = 1.00001$, the
quasiblack hole features show up. Namely, $B(r)\rightarrow \varepsilon $
for the whole interior region, $0\leq r < r_0$ and $1/A(r)\rightarrow
\varepsilon $ at $r = r_0$.

{\it The fluid quantities}:
The mass density $\rho_{\rm m}(r)$, the electric charge density
$\rho_{\rm e}(r)$, and the pressure $p(r)$ as a function of the
normalized radial coordinate $r/m$ are shown in the right panel of
Fig.~\ref{fig-a1pt2}, for the case $a=1.2$. As in the case of the potential
functions (see the left panel of Fig.~\ref{fig-a1pt2}), we plot the curves
in terms of the radial coordinate $r/m$ for five different values of
the normalized radius of the star $\beta = r_0/m$, namely, $\beta=
1.00001$ (solid line), $\beta = 1.2$ (space-dashed line), $\beta = 1.8$
(dashed line), $\beta = 2.5$ (dotted line), and $\beta =50$
(dash-dotted line).

\begin{figure}[h]
\begin{center}
\includegraphics[width=7.5cm,height=4.3cm]{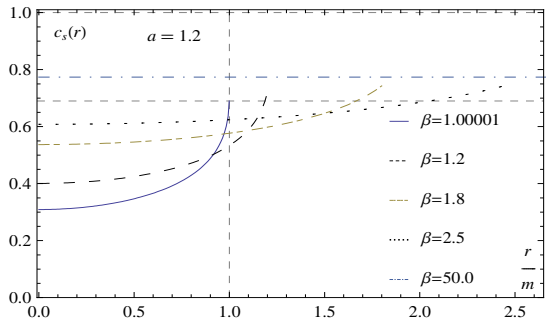}
\caption{The speed of sound $c_{\rm s}$ for $a =1.2$, as a function of the
normalized coordinate $r/m$, for five values of $\beta= r_0/m$ (from bottom
to top: $\beta=1.00001$, $\beta=1.2$, $\beta=1.8$, $\beta=2.5$, and
$\beta=50.0$). Notice that the speed of sound is smaller than the speed of
light throughout the star. }
\label{soundspeed_a1pt2}
\end{center}
\end{figure}

The case $\beta =50$ represents a very sparse star
and the fluid quantities are very small and are nearly constant in
the hole region inside the star, with the corresponding curves being the
lowest horizontal lines, almost coinciding with the horizontal axes line
(the bottom line) of the plot. In fact, such curves do not appear in
the graphs.
It is also seen that for the quasiblack hole case, represented in the
figures by the case $r_0/m = 1.00001$, the fluid quantities assume the
largest possible values. With the chosen value for the parameter $a$
($1\leq a\leq 4/3$), the fluid quantities are continuous decreasing
functions of the radial coordinate inside the star. The exterior region is
given by the Reissner-Nordstr\"om electrovacuum metric. The functions
$\rho_{\rm m}(r)$ and $\rho_{\rm e}(r)$ are truncated at $r=r_0$, resulting
in two discontinuous functions at that point which signal the jump into
vacuum. On the other hand, the function $p(r)$ is a monotonically
decreasing function of $r$, starting at $p(0)$ with the highest value and
reaching $p=0$ at $r=r_0$. We also see that the central pressure $p(0)$
increases with the compactness parameter $\beta$ of the star, analogous to
the behavior of the central pressure in white dwarfs, and contrary to the
behavior of the central pressure in main sequence stars.  One can also
verify numerically that the energy conditions listed in the previous
section are satisfied within the parameter space considered here.

Another final physical quantity worth of numerical analysis is the
speed of sound. The results confirm that the interesting solutions,
for which $c_{\rm s}$ is well-defined and the system preserves
causality, are those in which the parameter $a$ belongs to the
interval $1\leq a\leq 4/3$.  The curves for $c_{\rm s}(r)$ as a
function of $r/m$ for $a=1.2$ are shown in
Fig.~\ref{soundspeed_a1pt2}. The chosen values of the normalized
radius of the star are the same as for the other fluid quantities, as
in the right panel of Fig.~\ref{fig-a1pt2}. The horizontal dot-dashed
line is for a very sparse star, with $\beta =50$. The radial
dependence of the speed of sound is not seen because in this case the
radius of the star is too large. The other horizontal line, the dashed
line, indicates the maximum value of $c_{\rm s}(r)$ close to the
surface of the star in the quasiblack hole case, which is
approximately $0.7$. It also can be verified that for $a=4/3$ such a
maximum value is $c_{\rm s} \simeq 1$, for $r$ very close to $r_0$,
independently of how compressed or how sparse the star is. In fact,
the value of $c_{\rm s}(r_0)$ is the maximum value of the speed of
sound for all the stars within the parameter space considered in this
section. In turn, the minimum value of $c_{\rm s}(r)$ occurs at the
center of the star.\\

\subsubsection{The interval
$4/3 < a\lesssim 2.33$. Typical case: $a=1.7$}

In this section we show the main features of the metric and electric
functions and of the fluid quantities for a particular case of $a$ in the
interval $4/3 < a <2$. In this interval, all the potentials are
well-behaved functions of the radial coordinate for all the stars.
The fluid quantities are also smooth inside the star, and satisfy the
energy conditions (see Sec. \ref{sec-fluidqts}). However, the speed of
sound may reach values larger than the speed of light. Within this interval
we analyze the typical case $a=1.7$.

{\it The metric and the electric potentials}:
The general form of the curves for the metric potentials for $a=1.7$ is
nearly the same as for $a=1.2$ (see the left panels of
Figs.~\ref{fig-a1pt2} and \ref{fig-a1pt7}). This is true for all finite
values of $a$ in the intervals we investigated.  As seen from
Fig.~\ref{fig-a1pt7} the metric potentials are more sensible to the radius
to mass relation, $r_0/m$, than to the parameter $a$. The metric potential
$B(r)$ in the interior region decreases slightly with $a$, indicating that
the gravitational field strength increases with $a$. Similar small changes
are observed also in the metric function $A(r)$. The main changes, even
though small too, are in the electric potential $\phi(r)$ which, with our
choice of positive charge, increases with $a$. This indicates a noticeable
change in the central electric charge density.

\begin{figure}[h]
\begin{center} \vskip -.0cm
\includegraphics[width=7.cm,height=4.3cm]{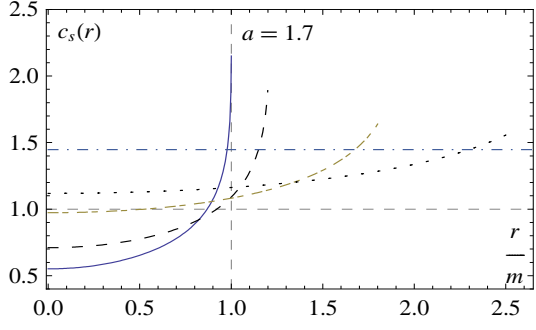}
\caption{The speed of sound $c_{\rm s}$, for $a=1.7$,
as a function of the normalized coordinate $r/m$, for five values of
$\beta= r_0/m$ in each graph (from bottom to top: $\beta=1.00001$,
$\beta=1.2$, $\beta=1.8$, $\beta=2.5$, and $\beta=50.0$). The case
$\beta=1$ is a quasiblack hole.}
\label{soundspeed_a1pt7}
\end{center}
\end{figure}

{\it The fluid quantities}:
A comparison between the right panels of Figs.~\ref{fig-a1pt2} and
\ref{fig-a1pt7} indicates that even though the fluid quantities strongly
depend on the parameter $a$, for the values of $a$ in the interval $4/3 < a
< 2$, the overall behavior of the density and pressure is similar to the
case $ 1\leq a\leq 4/3$. Namely, $\rho_{\rm m}(r)$, $\rho_{\rm e} (r)$, and
$p(r)$ are monotonically decreasing functions of the radial coordinate.
However, while the central value of the mass density remains almost the
same, the values of the pressure and of the charge density increase
substantially from $a=1.2$ to $a=1.7$.
In particular, the central pressure $p(r=0)$ increases by a factor of
approximately $4$.  Interestingly, even though the central charge
density increases with $a$, the total charge inside the star decreases
slowly with $a$, reaching zero as $a\to\infty$ for $r_0/m\geq 2$ as
discussed above. Moreover, in this interval, the energy conditions are
satisfied by all of the stars. On the other hand, the speed of sound
may be greater than the speed of light in some cases. This fact is
illustrated in Fig.~\ref{soundspeed_a1pt7}, where we see that the
curves representing $c_{\rm s}(r)$ for very compact stars, i.e., for
$\beta = r_0/m$ small, reach unity at some $r$ well inside the
star. It is also seen that, for sufficiently sparse stars, i.e., for
$\beta = r_0/m $ large enough compared to unity, the speed of sound is
larger than the speed of light everywhere inside the star.\\

\begin{widetext}
\vspace*{-.4cm}
\begin{figure}[h!]
\begin{center}
\includegraphics[width=7cm,height=11.3cm]{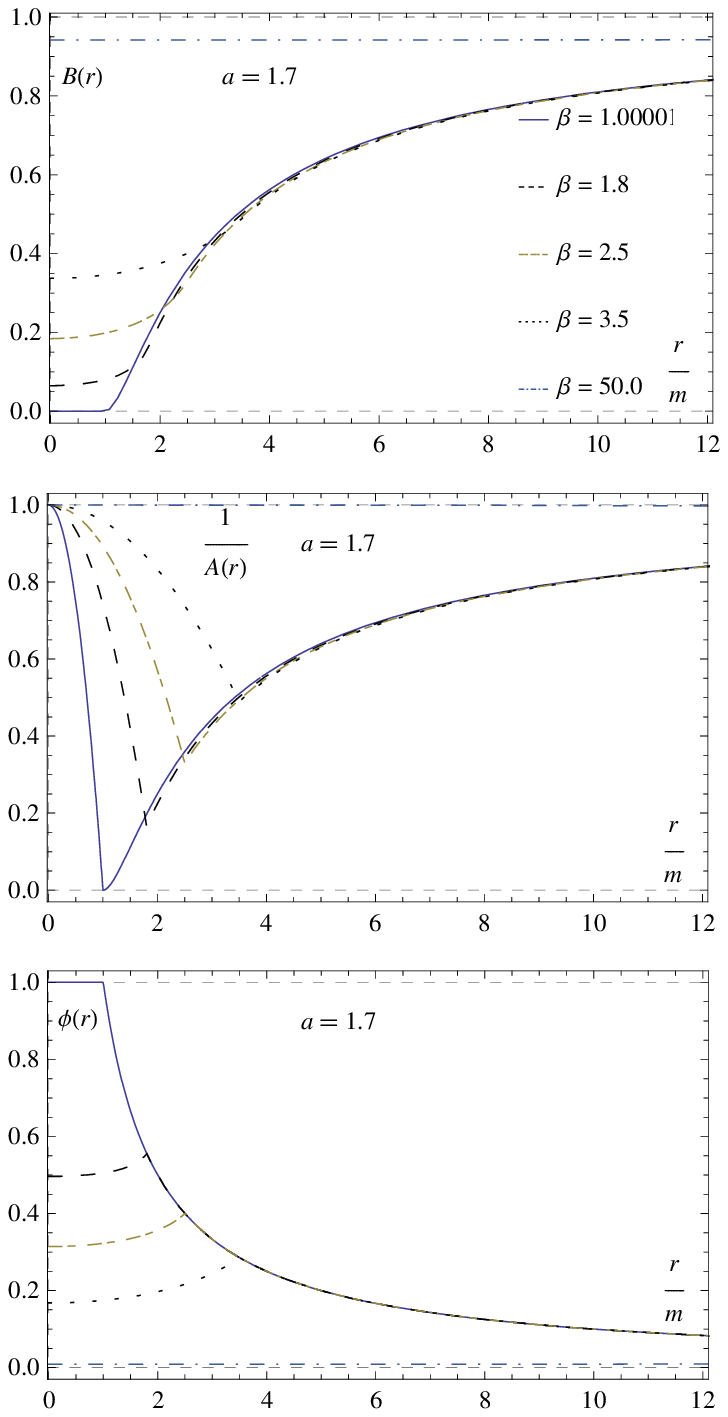}\hskip 1.3cm
\includegraphics[width=7cm,height=11.3cm]{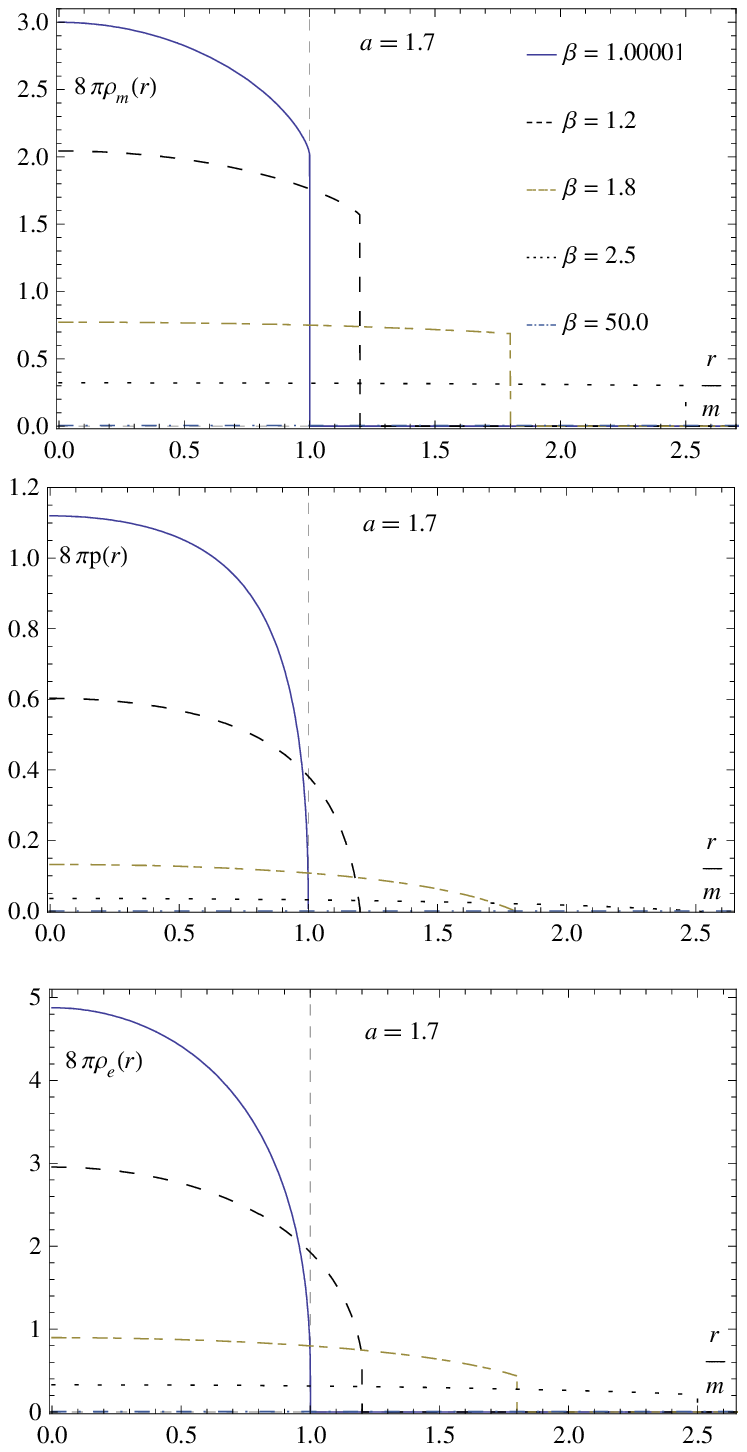}
\caption{ The same potentials (left panel) and the fluid quantities
(right panel) as in Fig.~\ref{fig-a1pt2}, but here for the case $a=1.7$.
As above, all quantities are plotted in terms of the normalized  radial
coordinate $r/m$, for five values of $\beta= r_0/m$ in each graph, and the
same conventions for the lines are used. Again, the case $\beta=1$, which
gives $q=m=r_0$, is a quasiblack hole. Notice that the pressure is larger
than in Fig.~\ref{fig-a1pt2}, but it is smaller than the energy density.}
\label{fig-a1pt7}
\end{center}
\end{figure}

\end{widetext}

\noindent

\subsubsection{The interval
$2.33<  a \leq 4$. Typical case: $a=3.0$}

Here we show the main features of the metric and electric functions
and of the fluid quantities for a particular case of $a$ in the
interval $2.33\leq a \leq 4$. In this interval, all the potentials are
well-behaved functions of the radial coordinate for all the stars.
The fluid also is smooth inside the star, and satisfies the energy
conditions (see Sec.  \ref{sec-fluidqts}). However, the speed of sound
reaches values larger than the speed of light and, for the most
compact stars, it diverges at some point inside the star.  Within this
interval we analyze the typical case $a=3.0$.

\begin{widetext}

\begin{figure}[htb]
\begin{center}
\includegraphics[width=7cm,height=11.3cm]{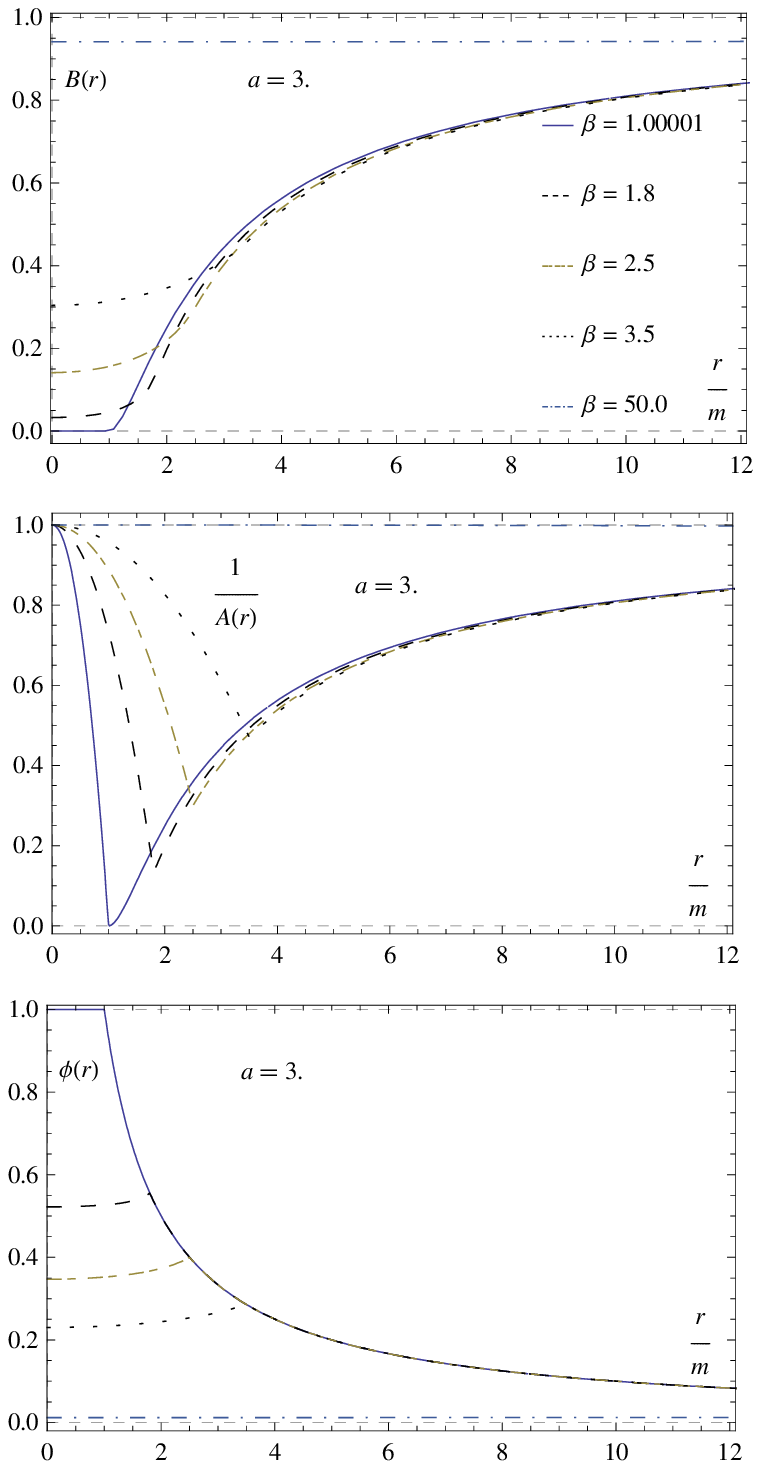}
\hskip 1.3cm
\includegraphics[width=7cm,height=11.3cm]{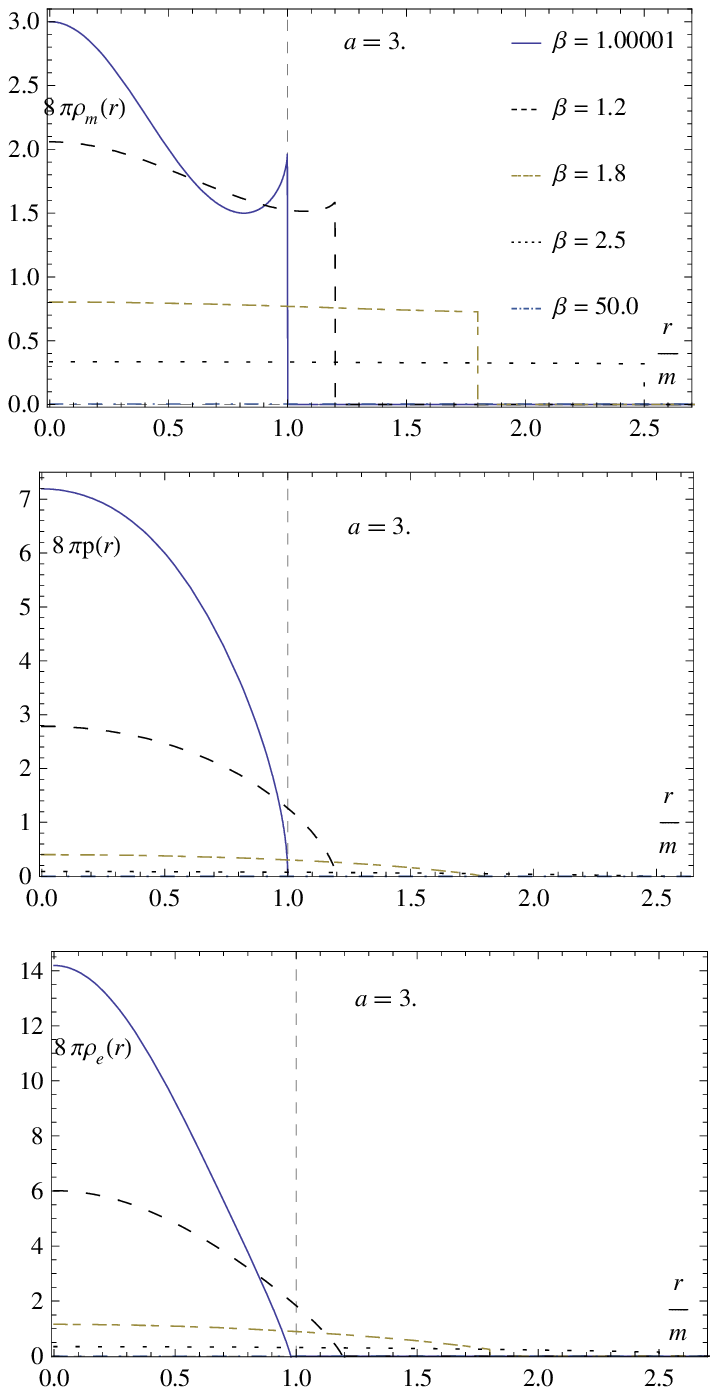}
\caption{ The same quantities as in Figs.~\ref{fig-a1pt2} and
\ref{fig-a1pt7} are plotted as a function of the normalized coordinate
$r/m$, but here for the case $a=3.0$. The same five values of $\beta=
r_0/m$, and the same conventions are used too. As above, the case
$\beta=1.2$ is not shown in the plots for the potentials, the case
$\beta =3.5$ being shown instead.  Notice that $\rho_{\rm m}(r)$ does
not decrease monotonically toward the surface $r=r_0$. Note also that
the central pressure is larger than the energy density.}
\label{fig-a3pt0}
\end{center}
\end{figure}

\end{widetext}

{\it Metric and electric potentials}:
The metric potentials $B(r)$ and $1/A(r)$, and the electric
potential $\phi(r)$ as a function of the normalized radial coordinate
$r/m$, for the case $a=3$, are shown in the left panel of
Fig.~\ref{fig-a3pt0}. The general features of the curves are the
same as in the case of the left panel of Figs.~\ref{fig-a1pt2} and
\ref{fig-a1pt7}, for which $a=1.2$ and $a =1.7$, respectively. By
comparing the three cases, the dependence of
these potentials upon the Guilfoyle parameter $a$
is now more clearly seen. As $a$ grows, the
central values of the metric potentials $B(r)$ and $1/A(r)$ diminish by a
small amount, and the electric potential $\phi(r)$ grows substantially
when compared to the case $a=1.2$. Of course, because of the
normalization used, these changes are not observed in the quasiblack hole
limit.

\begin{figure}[h!]
\begin{center}
\includegraphics[width=7.cm,height=4.0cm]{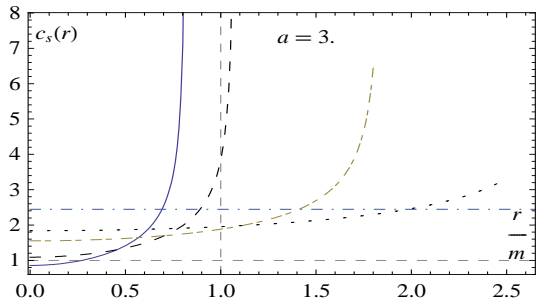}
\caption{The speed of sound $c_{\rm s}$, for $a=3.0$,
as a function of the normalized coordinate $r/m$, for five values of
$\beta= r_0/m$ in each graph (from bottom to top: $\beta=1.00001$,
$\beta=1.2$, $\beta=1.8$, $\beta=2.5$, and $\beta=50.0$). The case
$\beta=1$ is a quasiblack hole.}
\label{soundspeed_a3pt0}
\end{center}
\end{figure}

{\it The fluid quantities}:
The fact that the fluid quantities strongly depend on the parameter $a$ is
especially noticed for values of $a$ in the interval $2.33 \leq a < 4$, as
seen in the right panel of Fig.~\ref{fig-a3pt0}. An
interesting particularity is that, for sufficiently large $a$ and small
$\beta$, i.e., more compact stars, the energy
density $\rho_{\rm m} (r)$ is not a monotonically decreasing function of
$r$ anymore. As seen in that figure, the curve of $\rho_{\rm e} (r)$ for
the quasiblack hole case oscillates, attaining a minimum value for some
$r=r_m$ inside the star. This happens for all the sufficiently compact
stars. Also worthy of note is the fact that the central value of $\rho_{\rm
m}(r)$ does not depend on $a$. Meanwhile, the pressure $p(r)$ starts from a
relatively high value at the center of the star, and decreases very fast to
zero at the surface. As a consequence, even though the pressure is always
positive and a monotonically decreasing function of $r$, the speed of sound
increases very rapidly from the center, diverges at $r=r_m$, and becomes
undefined in the region $r> r_m$. For large $a$, the central value of the
electric charge density also becomes much larger than the central energy
density. 
Moreover, for $2.33\lesssim a < 4$, the dominant
energy condition is not satisfied by anyone of the stars. In fact,
excluding the central region of very compact stars, the speed of sound is
greater than the speed of light for all the stars. This fact is
illustrated in Fig.~\ref{soundspeed_a3pt0}, where we see that the curves
for $c_{\rm s}(r)$ are larger than unity for all $r$ inside the stars.

\section{Conclusions}
\label{sec-conclusion}

We have analyzed the class Ia of solutions provided by Guilfoyle
\cite{guilfoyle}.  Such spherically symmetric relativistic charged
fluid distributions are bounded by a surface of radius $r_0$.  The
interior region is filled with a fluid characterized by its mass and
charge densities and by a nonzero pressure.  The spacetime in the
exterior region is represented by the Reissner-Nordstr\"om metric.
These global solutions represent relativistic stars, i.e.,
relativistic cold charged spheres with pressure.  Besides the mass $m$
(or charge $q$, which are related to each other) and the radius of the
star $r_0$, this class of solutions is characterized by another free
parameter, the Guilfoyle parameter $a$. This parameter is related to
the pressure: for $a<1$ the stars are supported by tension; for $a=1$
the stars have no pressure (they are Bonnor stars), and for $a>1$ the
stars are supported by pressure.  The interval of the free parameter
$a$ can be fixed in such a way that the fluid satisfies the energy
conditions, and other physical requirements for a relativistic cold
star.  We have then studied relativistic stars within the interval
$1<a\leq 4$.  We have found that these cold stars show a mixed
behavior, in one instance they behave as main sequence stars in
another instance as white dwarfs.  Indeed, the mass to radius relation
of these cold stars is analogous to the behavior of the mass to radius
relation in main sequence stars and contrary to the mass to radius
relation in white dwarfs, whereas the central pressure of these cold
stars has an analogous behavior to the central pressure of white
dwarfs, and a contrary behavior to the central pressure of main
sequence stars.  We have also shown that, in the interval $1< a\leq
4$, the most compact configuration is a quasiblack hole with
pressure. Thus, quasiblack holes without pressure ($a=1$) studied
previously, as well as quasiblack hole with pressure ($1< a\leq 4$)
studied here, can be found within the context of general relativity.
As the most compact configuration is a cold star it can be called a
frozen star.

\begin{acknowledgments}
We thank Observat\'orio Nacional of Rio de Janeiro for hospitality
while part of the present work was being done. This work was partially
funded by Funda\c c\~ao para a Ci\^encia e Tecnologia (FCT) -
Portugal, through Projects Nos. CERN/FP/109276/2009
and PTDC/FIS/098962/2008.  VTZ thanks Funda\c
c\~ao de Amparo \`a Pesquisa do Estado de S\~ao Paulo (FAPESP)
 and Conselho Nacional de
Desenvolvimento Cient\'\i fico e Tecnol\'ogico of Brazil (CNPq) for
financial help.

\end{acknowledgments}

\end{document}